\documentclass[preprint,showpacs,preprintnumbers,amsmath,amssymb, showkeys]{revtex4}

\usepackage{array}
\usepackage{booktabs}
\usepackage{tabu}
\usepackage{dcolumn}
\usepackage{amsmath}
\usepackage{amsfonts}
\usepackage{amssymb}
\usepackage{graphicx}
\usepackage{subfigure}
\usepackage{graphicx}
\usepackage{dcolumn}
\usepackage{bm}

\begin{document}

\title{Renyi entropy and the holographic dark energy in flat space time}

\author{T. Golanbari}
  \email{t.golanbari@gmail.com; t.golanbari@uok.ac.ir}
\author{Kh. Saaidi}
  \email{ksaaidi@uok.ac.ir}
\author{Parvin. karimi}
  \email{Parvinkarimi452@gmail.com}

\affiliation{
Department of Physics, Faculty of Science, University of Kurdistan, Sanandaj, Iran.\\}
\date{\today}

\begin{abstract}
Based on Renyi entropy, we study the entropy corrected version of the holographic dark energy (HDE) model in apparent horizon of spatially flat Friedmann–Lemaître–Robertson–Walker (FLRW) universe. Applying the generalized entropy leads to the modified version of the Friedmann evolution equations which besides pressure-less matter and HDE, there is an extra term that is purely geometric. This extra term are assumed as another part of dark energy. We assume the universe is filled by non-interacting components of ideal fluids such as dark matter and holographic dark energy. The total dark energy, which is a combination of generalized HDE and geometric part, has a density parameter that approaches one by decreasing the redshift. Considering the total equation of state parameter and deceleration parameter of the universe indicates that the universe could stays in positive accelerated expansion phase that shows an agreement with observational data, only for the specific values of the constant $\zeta$.  \\
\end{abstract}

\maketitle

\section{Introduction}
\label{intro}
This is an acceptable belief that about ninety five percent of the universe is unknown which is a composition of two ambiguous components known as dark matter and dark energy \cite{Riess,Perlmutter,deBernardis,Perlmutter-a,Colless,Cole,Springel}. Dark matter is realized in astrophysical scale that contributes about $25\%$ of the universe. So far, its nature is unknown and we only have some candidate for describing this mystery such as axion, neutrino, primordial black holes. On the other hand, we have dark energy which fills about $70\%$ of the universe and works on cosmological scales and is responsible for the current positive accelerated expansion phase. \\
There are two general approaches to explain the reason of the present positive accelerated expansion phase of the universe. First, it is assumed that dark energy is a type of unknown and ambiguous fluid with negative pressure that is responsible for the present acceleration of the universe \cite{Wetterich,Caldwell,Armendariz,Cai,Wei,Nojiri,Ohta,Moradpour-0,Moradpour-0a,Nojiri-a,Bamba}; another way to describe the acceleration is applying modified theories of gravity where dark energy is explained in terms of geometrical effects \cite{Capozziello}. A new candidate for dark energy appeared when Cohen et. al. applied some hypothesis on the mutual relation between UV ($\Lambda$) and IR ($L$) cutoffs and the entropy of system, stating as $\rho_\Lambda \propto S / L^4$ where $\rho_\Lambda$ is the vacuum energy density \cite{Cohen}. In studying the thermodynamics of the black  hole \cite{Bekenstein,Hawking}, it is shown that the Bekenstein-Hawking entropy  bound $ S_{BH} \sim  M_{p} ^{2} L^{2} $, scales as the area $ A \sim L^{2}$ instead of the volume $ V \sim L^{3} $ ($ M_{p} $ is the reduced Plank mass, $ 8\pi G =1/M_{p}^{2} = 1$). Applying the idea to the cosmological studies led to a model for dark energy that is known as Holographic Dark Energy (HDE) \cite{Horava,Thomas,Hsu,Li,Odintsov,Odintsov1,Odintsov2,Odintsov3,Guberina,Myung}. The HDE has become an outstanding way to understand dark energy which has received huge attention \cite{Cohen,Thomas,Hsu,Li,Guberina,Ghaffari,Zadeh,Wang,Wang-a,Shen,Sheykhi,Zhang,Sheykhi-a,Setare,Sheykhi-b,Sheykhi-c,Karami,Agostino}. On the other hand, for the case which self-gravitation effects could be disregarded, the Bekenstein entropy bound $ S_{B} $ is EL where $ E = \rho_{\Lambda} L^{3} $ is the energy and L is the IR cutoff of system. Using $ S_{B} < S_{BH} $, one can see $ \rho_{\Lambda} \leq M_{p} ^{2} L^{-2} $. The HDE is given by defining a numerical constant c :
\begin{equation}\label{HDE}
\rho_{d}  = 3c^{2} M_{p} ^{2} L^{-2} .
\end{equation}
Observational data, which is obtained by constraining the HDE model, indicate that $ c = 0.818_{-0.097} ^{+0.113} $ \cite{Li-a} and
$ c = 0.815_{-0.139} ^{+0.179} $ \cite{Li-b} for flat and non-flat space time respectively.
More studies about HDE model and its features are done in \cite{Guberina,Hao,Wang,Wang-a,saaidi,saaidi-a,Aghamohammadi,karami-a,karami-b,karami-c,karami-d}. \\
The primary model of HDE that is based on attributing the Bekenstein-Hawking entropy to the cosmos horizon and assuming the Hubble horizon as the cutoff could not lead to a proper explanation for a spatially flat FLRW universe \cite{Horava,Thomas,Hsu,Li,Guberina}. One issue could be addressed to the fact that in this model both dark matter and dark energy evolve with the same function of scale factor \cite{Hsu,Li}. Although applying new cutoffs might solve the problem \cite{Hsu,Li}, even such cutoffs could not promise an stable model for dark energy as it becomes the dominant component \cite{Myung}. In general, the problem might be solved using different  methods such as: I) imposing other cutoffs than the Hubble cutoff, II) introducing possible interaction between the universe components, III) trying different entropies; or sometimes a combination of these methods \cite{Wang,Wang-a}. \\
According to the various research studies \cite{Masi,Touchette,Biro,Tsallis,Renyi,Tsallis-a,Abe,Abe-a,Majhi,Biro-a,Sayahian}, it seems that generalized entropy formalism, which are based on the power-law distribution of probability, is the prefer choice for systems that include long-range interactions, such as gravity. Due to the successes of the generalized entropy formalism in providing adequate explanation for dark energy, cosmologists have been motivated to apply the generalized entropy instead of the Bekenstein entropy as the horizon entropy \cite{Majhi,Sayahian,Komatsu,Moradpour,Moradpour-a,Moradpour-b,Abreu,Abreu-a,Barboza,Nunes}. Two recent generalized entropies could be named as Renyi and Tsallis \cite{Tsallis,Renyi,Tsallis-a,Odintsov4}, which widely have been utilized in studying different gravitational and cosmological phenomena \cite{Majhi,Biro-a,Sayahian,Bialas,Czinner,Komatsu,Moradpour,Moradpour-a,Moradpour-b,Abreu,Abreu-a,Barboza,Nunes}. Derivation of HDE density relies on the entropy-area relationship $S_{BH} = A/4$ where A is the area of horizon of black hole and $G = 1$. Therefore, by changing the entropy relation, one can find a new form of HDE. It is remarkable that by using a generalized form of entropy-area relation, the form of HDE and gravity model equations will be generalized. Therefore, we have a generalized Friedmann equation to describe the universe evolution. This fact motivates us to investigate the positive accelerating phase of the universe in the new form of HDE which is obtained based on Renyi entropy. It has been shown that, the Renyi entropy is given as \cite{Komatsu,Moradpour,Moradpour-a}
\begin{equation}\label{Renyientropy}
S_{R}  = \frac{1}{\zeta} \ln ( 1+ \zeta S_{T} ) ,
\end{equation}
where $ S_{T} $ is the Tsallis entropy and $\zeta$ is a small real constant. It has been explored that the Tsallis and Bekenstein is equal \cite{Komatsu,Moradpour,Moradpour-a,Czinner}. Then,  Eq.(\ref{Renyientropy}) is reduced to:
\begin{equation}\label{RenyiS}
S_{R}  = \frac{1}{\zeta} \ln ( 1+ \zeta \frac{A}{4} ) ,
\end{equation}
in which by approaching the constant $\zeta$ to zero, i.e. $ \zeta \rightarrow 0 $, the Renyi entropy reduces to $ A/4 $. \\
During the present work, we are going to consider the HDE by applying the Renyi entropy. It is assumed that the universe is described by a flat FLRW metric, filled with non-interacting components. It should be mentioned that the same topic was studied in EPJC (2018) \cite{Moradpour-c}, however the problem is that even by applying a generalized entropy, by mistake the authors performed the work by considering the standard Friedmann equation which certainly comes to unreliable results. The point is that due to applying a generalized entropy, a generalized Friedmann equation is resulted which includes an entropy corrected HDE plus an extra term which is purely geometric. Then, the situation become more complicated because there is an extra density parameter related to the constant $\zeta$ which is the result of the Renyi entropy.

\section{The model}\label{sec:1}

As we mentioned before, Renyi entropy is a generalize version of Bekenstein entropy so that the gravity and then Friedmann equations based on it should be generalized. Therefore, firstly, we are about to find the modified version of Friedmann equations based on the Renyi entropy. We propose the spatially flat FLRW metric as
\begin{equation}\label{FLRW}
dS^2  = dt^2 - a^2(t) \; d\mathbf{x}.d\mathbf{x},
\end{equation}
where $ a(t) $ is the scale factor of the universe. The apparent horizon in the flat FLRW universe is given as \cite{Poisson,Hayward,Gong}.
\begin{equation}\label{apparent-horizon}
r  = \frac{1}{H} ,
\end{equation}
and the Howking temperature on the horizon is obtain as \cite{Jacobson}
\begin{equation}\label{HowkingTemp}
T = \frac{1}{2 \pi r} = \frac{H}{2 \pi} .
\end{equation}
To derive the modified Friedmann equation based on Renyi entropy, we use the Clausius relation \cite{Cai-a}.
\begin{equation}\label{Clausivs}
-dE = T dS ;
\end{equation}
Here $ -dE $ is the energy crossing from a fixed horizon in the infinitesimal time interval. According to \cite{Cai-b}, one can obtain :
\begin{equation}\label{dE}
dE = - 4 \pi r^{3} ( p + \rho ) H dt ,
\end{equation}
where, $p$ and $ \rho $ are the pressure and energy density of ideal fluid, respectively. We assume the dark energy component in the universe is given by HDE and total energy inside the universe satisfy the following relation:
\begin{equation}\label{TotalEnergy}
\dot{\rho} + 3 H ( p + \rho ) = 0 ,
\end{equation}
Where $ \rho = \rho_{m} + \rho_{d} $ and $p=p_m+p_d$ are respectively the total energy density and total pressure of the universe respectively.
Using (\ref{RenyiS}), (\ref{apparent-horizon}), (\ref{HowkingTemp}), (\ref{Clausivs}), (\ref{dE}), and (\ref{TotalEnergy}), we have
\begin{eqnarray}
H^{2} - \zeta \pi \ln ( H^{2} + \zeta \pi ) &=&\frac{8 \pi }{3} (\rho_{m} + \rho_{d} ) , \nonumber \\
 H^{2} - \zeta \pi \ln ( H^{2} + \zeta \pi ) + \frac{\frac{2 \dot{H}}{3}}{1 +\frac{\zeta \pi}{H^{2}}}&=& - \frac{8\pi }{3} p_{d} , \label{FriedmannEq}
\end{eqnarray}
where $ \rho_{m} $ and $ \rho_{d} $ are energy density for matter and dark energy respectively and $ p_{d} $ is the pressure
of HDE. It is obviously seen that for $ \zeta = 0 $, these equations
reduce to the ordinary Friedmann equation. As we mentioned before, we assume the vacuum energy density has the role of dark energy, namely we suppose, $ \Lambda = \rho_{\Lambda} = \rho_{D} $ which is given by holographic energy in this model i.e. $ r_{D} = r_{d} $ therefore, by
utilizing Eqs. (\ref{apparent-horizon}), (\ref{HowkingTemp}), $ A = 4 \pi r^{2} $ and $ \rho_{\small{d}}
dV \propto T dS $ one can obtain:
\begin{equation}\label{rho_d-propto}
\rho_{d} \propto \frac{H^{2}}{4 \pi ( 1 + \frac{\zeta \pi }{H^{2}})} ,
\end{equation}
Defining a constant, we can write Eq. (\ref{rho_d-propto}) as
\begin{equation}\label{rho_d-propto1}
\rho_{d} = \frac{3 c^{2} H^{2}}{8 \pi ( 1 + \frac{\zeta \pi
}{H^{2}} )} ,
\end{equation}
Note that for $ G = 1 $ , $ M_{p} ^{-2} = 8 \pi $ and by inserting $
\zeta = 0 $ Eq. (\ref{rho_d-propto1}) reduce to Eq. (\ref{HDE}) and Eq.(\ref{FriedmannEq}) reduces to ordinary Friedmann equation. This fact shows that our modified
version of holographic dark energy is combatable with the ordinary one Eq. (\ref{HDE}).


In this work, we assume that there is no interaction between main component of the universe, and the EoS could be written as
\begin{eqnarray}
\dot{\rho}_{d} + 3H ( \rho_{d} + p_{d} ) &=& 0, \label{dot-rho_d}  \\
\dot{\rho}_{m} + 3H \rho_{m} &=& 0 \label{dot-rho_m}
\end{eqnarray}
From Eqs. (\ref{dot-rho_d}) and (\ref{dot-rho_m}), we have
\begin{eqnarray}
p_{d} &=& -  \rho_{d} - \frac{\dot{\rho}_{d}}{3 H}, \label{p_d},  \\
\rho_{m}&=& \rho_0 (1 + z )^{3}, \label{rho_m}
\end{eqnarray}
where $ \rho_{0} $ is a constant and $z$ is the redshift.\\
Using the following definitions for density parameters
\begin{equation}\label{Omega_m-d}
\Omega_{m} = \frac{8 \pi }{3 H^{2}} \rho_{m}  \quad , \quad
\Omega_{d} = \frac{8 \pi }{3 H^{2}} \rho_{d},
\end{equation}
and
\begin{equation}\label{Omega_zeta}
\Omega_{\zeta} = \frac{\zeta \pi \ln(H^{2} + \zeta \pi )}{ H^{2}},
\end{equation}
one can rewrite the first generalized Friedmann equation (\ref{FriedmannEq}) as
\begin{equation}\label{Omega_m-d-zeta}
1 = \Omega_{m} + \Omega_{d} +\Omega_{\zeta}
\end{equation}

Inserting Eqs. (\ref{p_d}) and (\ref{rho_d-propto1}) in Eq. (\ref{FriedmannEq}), one obtains
\begin{equation}\label{E2-z1}
E^{2} (z) - \frac{\zeta \pi}{H_{0} ^{2}} \ln ( H^{2} +\zeta \pi ) = \Omega_0 ( 1+z )^{3} + \frac{c^2}{( 1 + {\zeta \pi \over E^2(z)
H_0^2} )} E^{2} (z) ,
\end{equation}
where $ H ( z ) = E ( z ) H_{0} $ and $ H_{0} $ is the present Hubble parameter, namely   $H_0= H( z= 0) $. Than
\begin{equation}\label{Omega0-z}
\Omega_0 ( z = 0 ) =  \frac{8 \pi \rho_{0} }{3 H_0^2} .
\end{equation}
In obtaining Eq. (\ref{E2-z1}), it is assumed that $E(z)=1$ for $z=0$, and this assumption leads to the following expression for the constant $c$, i.e.
\begin{equation}\label{c}
c^2= (1 + \frac{\zeta \pi}{H_0^2})\left( 1 - \Omega_0 -\frac{\zeta \pi}{H_0^2} \ln ( H_0^{2} +\zeta \pi )\right) .
\end{equation}
On the other hand, from Eq. (\ref{Omega_zeta}) and (\ref{Omega_m-d-zeta}), one can obtain the following relation,
\begin{equation}\label{E2-z2}
E^{2} (z) {\bigg(} 1 - {\Omega_{d}} - {\Omega_{m}} {\bigg)} =
(\frac{{\zeta }{\pi}}{ H_{0}^{2}}) \ln ( H^{2} +{\zeta }{\pi} ) ,
\end{equation}
and by substituting Eqs. (\ref{E2-z1}) in (\ref{E2-z2}), one yields
\begin{equation}\label{E2-z3}
E^{2} (z) = \frac{\Omega_0 ( 1 + z )^3}{\Omega_m + \Omega_{d}} +
\frac{c^2}{(\Omega _{d} + \Omega _{m})(1 + \frac{\zeta \pi}{E^2 (z) H_0^2})} E^2(z) ,
\end{equation}
From Eq.(\ref{E2-z3}), the parameter $E(z)$ could not be derived analytically which this is a direct result of the modified Friedmann equation.
In \textit{EPJC 78, 89 (2018)},
where the authors derived an analytical solution for $E(z)$, the originally form of the Friedmann equation is utilized, which according to the fact that we are working with generalized entropy, is an incorrect conclusion, and this output affects all their next results.


\section{The cosmological parameters behavior}

\begin{figure}
\centering
          \includegraphics[width=7cm]{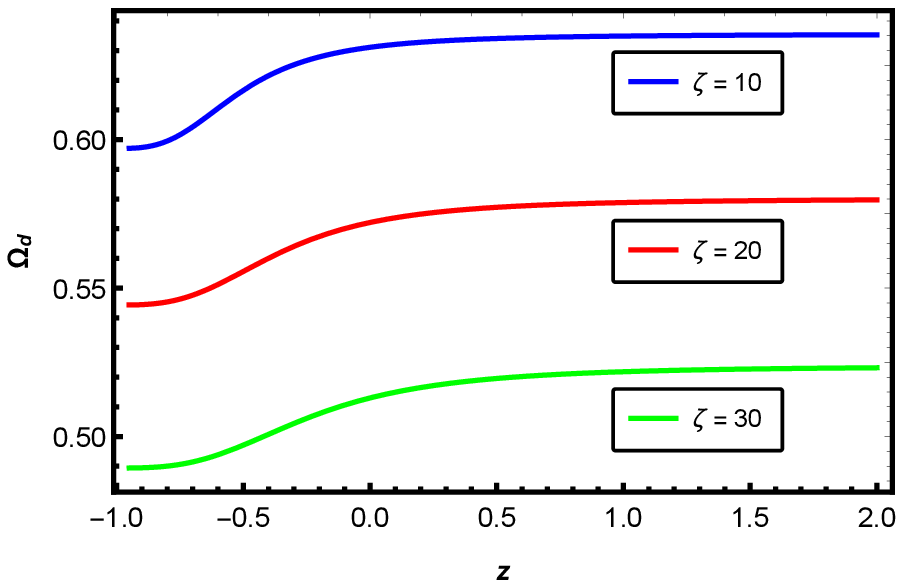}
          \includegraphics[width=7cm]{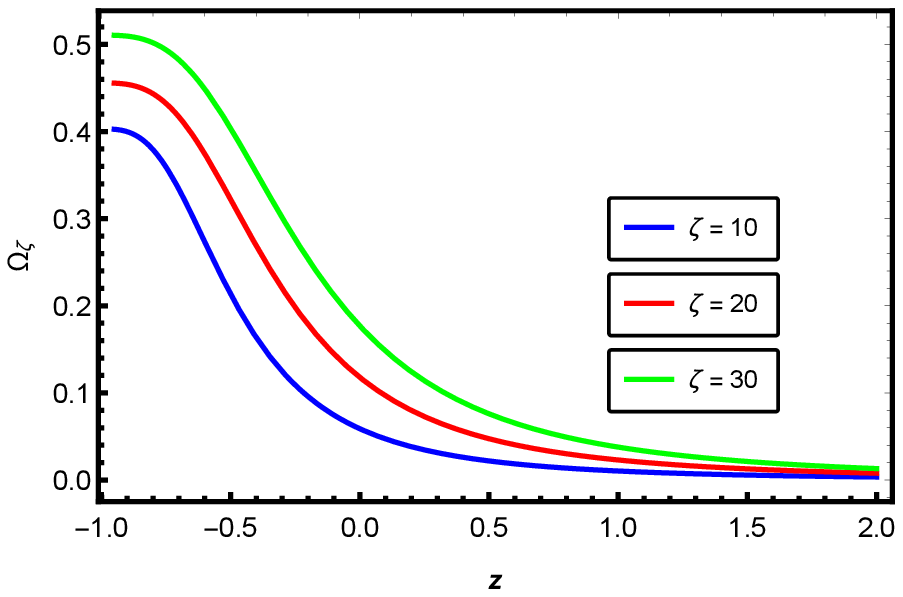}
          \includegraphics[width=10cm]{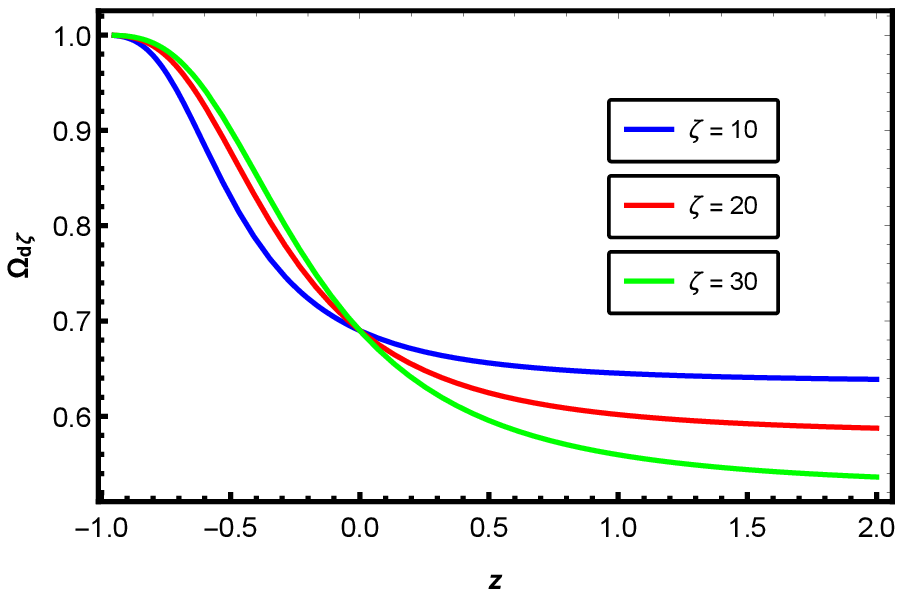}
\caption{The density parameters $ \Omega_{d}, \Omega_{\zeta}$ and $ \Omega_{\zeta d} $ versus z for different choice of $ \zeta $. Here, $ \Omega_{m} = 0.31 $ and $ H_{0}= 67  $(Km/s)/Mpc \cite{Planck2018}.}
\label{omega}
\end{figure}

To consider some of cosmological parameters, such as density parameter and equation of state parameter, we are going to apply the numerical solution for $E(z)$, and by using this solution, the cosmological parameters and their behavior will be plotted and studied versus the redshift, $z$.
The obtained solution for $E(z)$ is used to determine the behavior of the density parameter of the total dark energy and its components. In Fig. \ref{omega}, the density parameters $\Omega_d$ and $\Omega_\zeta$ have been plotted versus redshift $z$ for different choices of $ \zeta $, where the constants $\Omega_{m} = 0.31 $ and $ H_{0}= 67 \; {\rm km \; s^{-1} Mpc^{-1}}$ are taken based on the last observational data of Planck \cite{Planck2018}. Note that for these values of the parameters $\Omega_m$ and $H_0$ for the present time, the constant $c$ could be read from Eq.(\ref{c}) which remains only as a function of $\zeta$. For $\zeta=10,20$ and $30$ that we choose for our work, the constant $c$ stands between $c=0.79$ to $0.82$, and this range of $c$ is in good agreement with the value of the same parameter that has been obtained in \cite{Miao-Li} for the holographic model. \\

From Fig.\ref{omega} it is seen that the density parameter of the HDE is higher for bigger values of the redshift, and by passing time it decreases. The effect of the constant $\zeta$ on the density parameter $\Omega_d$ is shown in the figure in which by increasing the constant, the density parameter takes smaller values. On the other hand, the density parameter of the geometric part has a different behavior in which it is small for high values of the redshift, and it increases by reduction of the redshift. Also, for smaller values of the constant $\zeta$ the density parameter $\Omega_\zeta$ receives smaller values which is oppose to the behavior of $\Omega_d$. The density parameter of the total dark energy, $\Omega_{d \zeta}$, which is a combination of  the HDE and geometric energy is also depicted in Fig. \ref{omega}, which indicates that the density parameter of the total dark energy of the Universe grows by passing time and approaches one for negative values of the redshift, stating that the universe will be dominated by dark energy.  \\

In addition using Eqs. (\ref{FriedmannEq}) and (\ref{p_d}), one can reach to the following equation for $\dot{H}$,
\begin{equation}\label{dotH}
\dot{H} = \frac{4 \pi c^2 \rho_{m}}{ \Omega_{d} (2c^2 - 1  - \Omega_{d}) } ,
\end{equation}
in which to reach the above expression, the following relation is applied
\begin{equation}\label{rho-prime _d}
\rho^\prime _{d} = \frac{d\rho_{d}}{dH } = \frac{2 \rho_{d}}{c^2 H} ( 2c^2 -  \Omega_{d} ).
\end{equation}
Substituting Eqs. (\ref{dotH}) and (\ref{rho-prime _d}) in Eq. (\ref{p_d}), we have
\begin{equation}\label{p_d2}
p_{d} = - \left[ 1 + (\frac{2c^2 -  \Omega_{d}}{2c^2 - 1  - \Omega_{d}}) (\frac{\Omega_{m}}{\Omega_{d}}) \right] \; \rho_{d}.
\end{equation}

\begin{figure}
\centering
          \includegraphics[width=7cm]{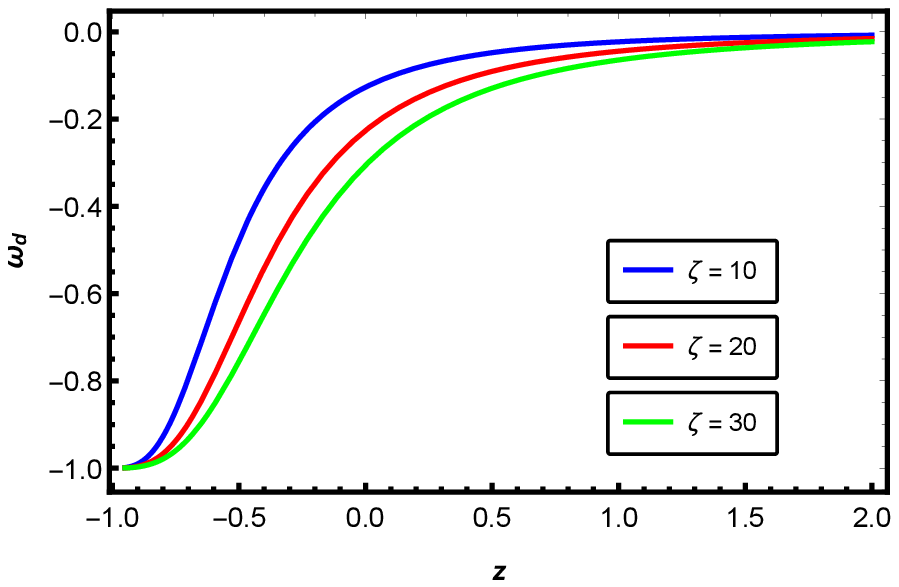}
          \includegraphics[width=7cm]{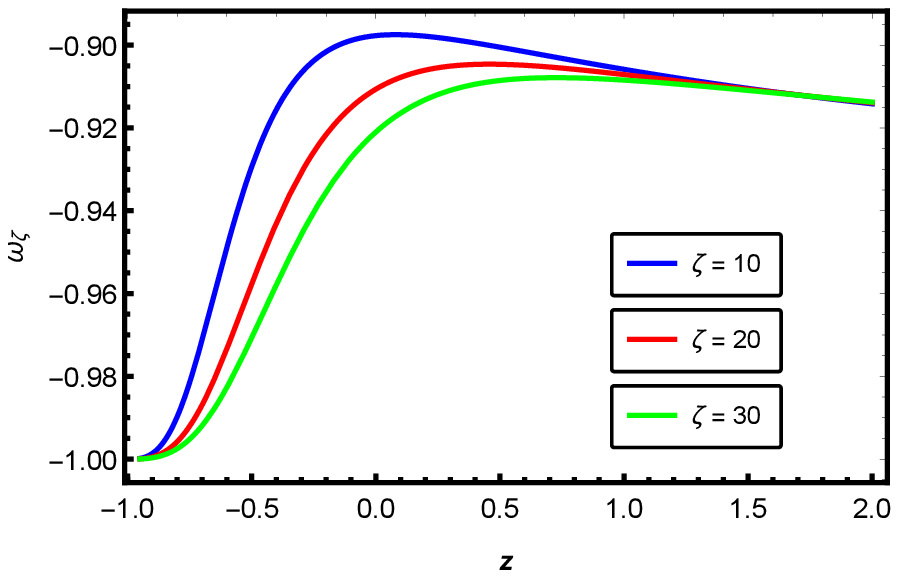}
          \includegraphics[width=10cm]{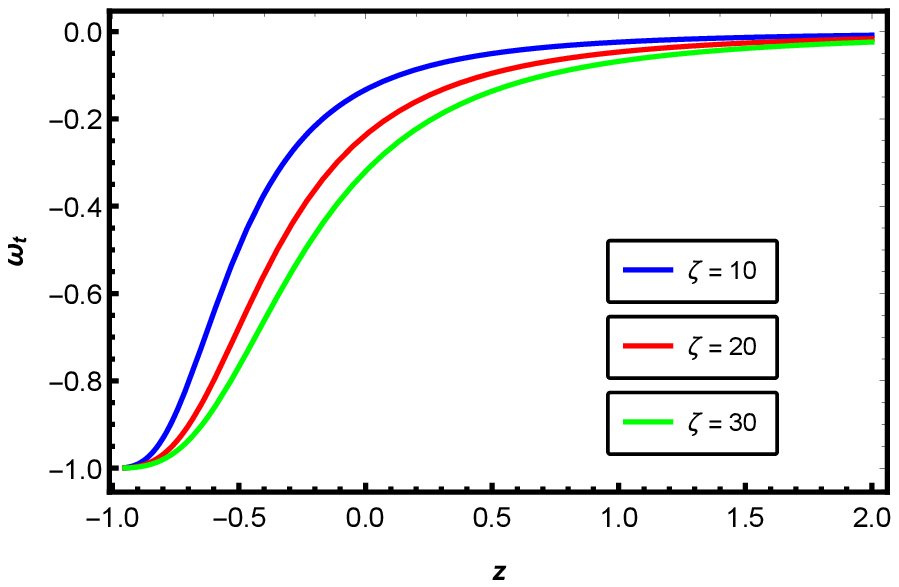}
\caption{The  EoS  $ w_{d}, w_{\zeta d}, w_{t} $ versus z for different choice of $ \zeta $. Here, $ \Omega_{m} = 0.31 $ and $ H_{0}= 67  $(Km/s)/Mpc \cite{Planck2018}.}
\label{w-d}
\end{figure}

By defining the equation of state (EoS) parameter of HDE as $\omega_{d} = p_{d} / \rho_{d}$, the following relation is concluded for $\omega_{d}$

\begin{equation}\label{omega_d}
\omega_{d} = -1 - (\frac{2c^2 - \Omega_{d}}{2c^2 - 1  - \Omega_{d}}) (\frac{\Omega_{m}}{\Omega_{d}})
\end{equation}
With attention to this fact that the modified term in the Friedmann equation (\ref{FriedmannEq}) is considered as a fluid, one could assume it as a perfect fluid and define an EoS. Following the same process as above, the EoS parameter for this part is read as
\begin{equation}\label{omega_d}
\omega_{\zeta} = -1 - (\frac{\zeta \pi}{H^2(2c^2 - 1  - \Omega_{d})}) (\frac{\Omega_{m}}{\Omega_{\zeta}}).
\end{equation}
Also, we can characterize the total EoS parameter as $ \omega_t \equiv (p_\zeta + p_d)/(\rho_{\zeta} + \rho_{d} + \rho_{m}) $, then
\begin{equation}\label{omega_d}
\omega_{t} = -\left( 1 + \frac{4}{3}\frac{\dot{H}}{H^2}-\frac{2\zeta \pi}{3c^2}\frac{\dot{H}}{H^4}\right)\Omega_{d} - \Omega_{\zeta} + \frac{2}{3}\frac{\dot{H}}{H^2}\frac{\Omega_{d}^2}{c^2}
\end{equation}
In Fig. \ref{w-d}, the EoS parameters of two components of dark energy, $ \omega_{d}$ and  $\omega_{\zeta} $, and also the EoS parameter of the total dark energy $ \omega_{t} $, have been plotted in terms of $z$ for different choices of $ \zeta $, where $ \Omega_{m} = 0.31 $, $ H_{0}= 67  {\rm (km \ s^{-1} Mpc^{-1})}$.

The parameter $\omega_d$ corresponds to the HDE and $\omega_\zeta$ is related to the geometric part of dark energy. Both parameters are negative which indicates both components possess negative pressure, which in turn states that the geometric component could also be taken as a type of dark energy. In addition, the parameter $\omega_d$ approaches $-1$ by reduction of the redshift, however the parameter $\omega_\zeta$ increases a little at first, then it decreases and tends to $-1$ as well. The total EoS parameter is depicted versus the redshift $z$ in Fig.\ref{w-d} for different values of the constant $\zeta$. The figures shows that for the present time, i.e. $z=0$, the total EoS parameter in below $-1/3$ for $\zeta=30$ which implies on a universe with accelerated expansion phase. For two other choices of $\zeta$, $\omega_t$ is negatives but it is higher than $-1/3$ and could not describe the positive accelerated phase of the universe for the present time. Another point is that, Fig.\ref{w-d} determines that the parameter $\omega_t$ ia approaching $-1$ which states that our universe tends to reach a de Sitter universe.   \\

Moreover, deceleration parameter, q, is find out as
\begin{equation}\label{q-deceleration1}
q = - 1 - \frac{\dot{H}}{ H^{2}} ,
\end{equation}

\begin{figure}
\centering
          \includegraphics[width=10cm]{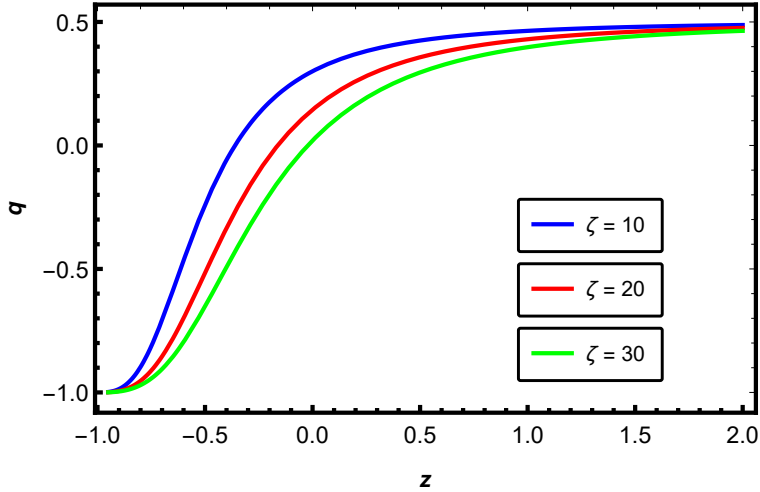}
\caption{The deceleration parameter $ q $ versus z for different choice of $ \zeta $. Here, $ \Omega_{m} = 0.31 $ and $ H_{0}= 67 (Km/s)/Mpc $.}
\label{q}
\end{figure}

Using Eq. (\ref{dotH}), we reach
\begin{equation}\label{q-deceleration2}
q = - 1 - \frac{\Omega_{m}}{\Omega_{d}}\left( \frac{3c^2}{2(2c^2 - 1  - \Omega_{d})} \right)
\end{equation}
In Fig. \ref{q}, deceleration parameter, q, have been plotted versus z for different choices of $ \zeta $, where $ \Omega_{m} = 0.31 $ and $ H_{0}= 67  $(Km/s)/Mpc.

For all cases, there is a phase transition in which the universe exits from a deceleration phase and enters to a positive accelerated expansion phase. It is clear that when the universe is dominated by matter, the deceleration parameter is about $1/2$, and by passing time, when the universe would be overwhelmed by dark energy, the deceleration parameter reaches negative values. However, the results depends on the values of the constant $\zeta$ so that for $\zeta=10$ and $20$ the deceleration parameter is positive for the present time ($z=0$) that shows a negative accelerated phase for the universe that is in clear contraction with observation. However, for $\zeta=30$, the deceleration parameter $q$ is negative for the present time. This result is in complete agreement with our result that has been obtained about the total EoS parameter $\omega_t$.  \\

\begin{figure}
\centering
          \includegraphics[width=7cm]{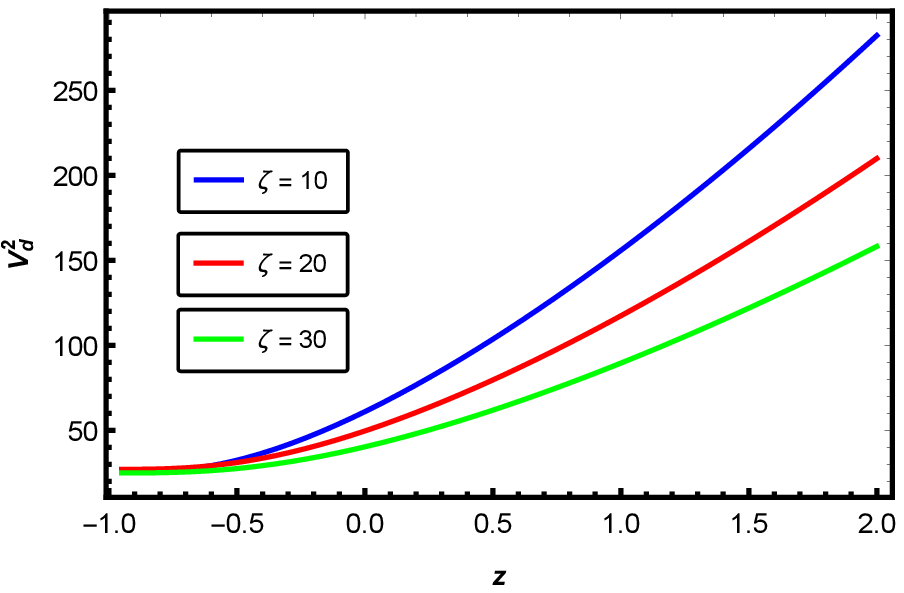}
          \includegraphics[width=7cm]{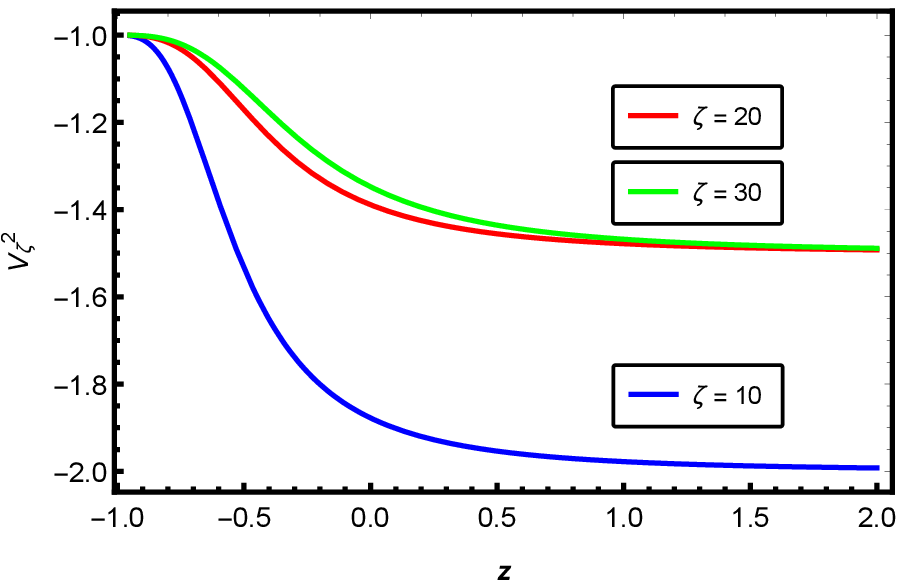}
          \includegraphics[width=10cm]{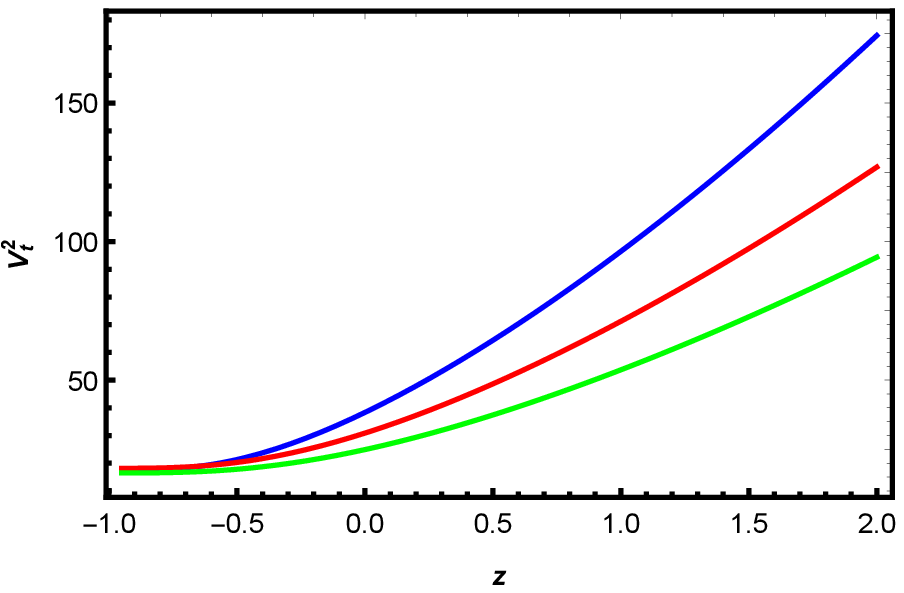}
\caption{The squared sound speed $ v_{d}^{2}, v_{\zeta}^{2}, v_{t}^{2} $ versus z for different choice of $ \zeta $. Here, $ \Omega_{m} = 0.31 $ and $ H_{0}= 67 (Km/s)/Mpc $ \cite{Planck2018}.}
\label{SoundSpeed-d}
\end{figure}

Finally, we consider the stability of the model against perturbation. For this goal we explore the squared sound speed of the components of dark energy and the total fluid of the universe, which are defined as

\begin{equation}\label{v2sd}
v_{i}^2 = {dp_i \over d \rho_i} = ({\rho_i \over \rho_i^{\prime}}) \omega_{i}^{\prime} + \omega_{i}
\end{equation}
where $i$ stands for $i=d,\zeta, t$ and prime indicates derivative with respect to $H$. Note that the sign of squared sound speed ( $v_i^2 >0 $ or $v_i^2<0$) determines the stability or unstability of each component against perturbation respectively. Fig.\ref{SoundSpeed-d} displays the squared sound speed for generalized HDE, geometric dark energy and the model. It could be realized that the sound speed of generalized HDE is positive which indicates that this part of dark energy is stable. The most important case is the stability of the total fluid of the universe  which according to the Fig.\ref{SoundSpeed-d} is stable and this result in turn implies on the stability of the model.



\section{Conclusion}
 There are many candidates for dark energy to explain the positive accelerated expansion phase of the universe, in which HDE is one of them that is known as a promising model to explain the nature of this ambiguous fluid. HDE model, as one of those candidates of dark energy, is based on the entropy-area relation. Since the entropy-area relation depends on the gravity theory, generalizing one of them, i.e. gravity or entropy, will modified the other one.  Based on this fact, we found out a modified Friedmann evolution equation by using a type of generalized entropy which is known as Renyi entropy. In addition by identification of IR cut-off with apparent horizon, $ L = H^{-1} $, and exploring the Bekenstein entropy as the Tsallis entropy, and by using the Renyi entropy we obtained the modified version of HDE.  Besides ordinary matter and generalized HDE, the modified friedmann equations contains an extra term which are purely geometric. This term possesses a negative pressure and because of this fact, it was taken as a type of dark energy. Then, it was assumed that the total dark energy contains two parts, one the generalized HDE and the other is geometric dark energy. \\
In the work, the effects of generalized version of entropy on various properties of holographic dark energy, cosmological and gravitational phenomena were studied. Considering the density parameter of dark energy shows that the total density parameter of dark energy increases by passing time, and approaches to one in future and dominates the universe. The equation of state parameters of dark energy and its components were obtained and plotted, and the results shows that EoS parameters of generalized HDE and the geometric dark energy tends to $-1$ by decreasing redshift, which indicates that the EoS parameter of the total dark energy also approaches to the bigger negative values by passing time, which indicates that the universe is going to a positive accelerated expansion phase. However, the result depends on the values of the constant $\zeta$, and only for some specific values of this constant, the current universe stands in positive acceleration. For instance, for $\zeta=10$ or $20$ the universe is in deceleration phase in the present time, however for $\zeta=30$ the acceleration of the universe at the present time is positive. This result is in complete agreement with the behavior of the deceleration parameter which was considered in next step. The behavior of the deceleration parameter was illustrated in Fig.\ref{q}, which shows that there is a phase transition and the universe goes from deceleration phase (matter dominant era) to a positive accelerated expansion phase (dark energy dominant era). It seems that the current positive acceleration phase of the universe is acceptable for higher values of the constant $\zeta$.


\section{Acknowledgments}
The work of T. G. has been supported financially by "Vice Chancellorship of Research and Technology, University of Kurdistan" under research Project No. 98/11/2724.


\begin{thebibliography}{}

\bibitem{Riess} A. G. Riess et al., Astron. J. \textbf{116}, 1009 (1998).

\bibitem{Perlmutter}S. Perlmutter et al., Astrophys. J. \textbf{517}, 565 (1999).

\bibitem{deBernardis} P. de Bernardis et al., Nature \textbf{404}, 955 (2000).

\bibitem{Perlmutter-a} S. Perlmutter et al., Astrophys. J. \textbf{598}, 102 (2003).

\bibitem{Colless} M. Colless et al., Mon. Not. R. Astron. Soc. \textbf{328}, 1039 (2001).

\bibitem{Cole} S. Cole et al., Mon. Not. R. Astron. Soc. \textbf{362}, 505 (2005).

\bibitem{Springel} V. Springel, C.S. Frenk, S.M.D.White, Nature (London) \textbf{440}, 1137 (2006).



\bibitem{Wetterich} C. Wetterich, Nucl. Phys. B \textbf{302}, 668 (1988).

\bibitem{Caldwell} R. R. Caldwell, M. Kamionkowski, N. N. Weinberg, Phys. Rev. Lett. \textbf{91}, 071301 (2003).

\bibitem{Armendariz} C. Armendariz-Picon, V. F. Mukhanov, P. J. Steinhardt, Phys. Rev. Lett. \textbf{85}, 4438 (2000).

\bibitem{Cai} R. G. Cai, Phys. Lett. B \textbf{657}, 228 (2007).

\bibitem{Wei} H. Wei, R. G. Cai, Phys. Lett. B \textbf{660}, 113 (2008).

\bibitem{Nojiri} S. Nojiri, Sergei D. Odintsov, Phys. Rev. D \textbf{72}, 023003 (2005).

\bibitem{Ohta} N. Ohta, Phys. Lett. B \textbf{695}, 41 (2011).

\bibitem{Moradpour-0} H. Moradpour, A. Abri, H. Ebadi, Int. J. Mod. Phys. D \textbf{25}, 1650014 (2016).

\bibitem{Moradpour-0a} H. Moradpour, R. C. Nunes, E. M. C. Abreu, J. A. Neto, Mod. Phys. Lett. A \textbf{32}, 1750078 (2017).

\bibitem{Nojiri-a} S. Nojiri, S. D. Odintsov, Phys. Lett. B \textbf{639}, 144 (2006).

\bibitem{Bamba} K. Bamba, S. Capozziello, S. Nojiri, S. D. Odintsov, Astrophys. Space Sci. \textbf{342}, 155 (2012).


\bibitem{Capozziello} S. Capozziello, V. Faraoni, \textit{Beyond Einstein Gravity} (Springer, NY, 2011).


\bibitem{Cohen} A. G. Cohen, D.B. Kaplan, A.E. Nelson, Phys. Rev. Lett. \textbf{82}, 4971 (1999).


\bibitem{Bekenstein} J. D. Bekenstein, Phys. Rev. D \textbf{7}, 2333 (1973);  Phys. Rev. D \textbf{9}, 3292 (1974); Phys. Rev. D \textbf{23}, 287 (1981); Phys. Rev. D \textbf{49}, 1912 (1994).

\bibitem{Hawking} S. W. Hawking, Commun. Math. Phys. \textbf{43}, 199 (1975); Phys. Rev. D \textbf{13}, 191 (1976).


\bibitem{Horava} P. Horava, D. Minic, Phys. Rev. Lett. \textbf{85}, 1610 (2000) .

\bibitem{Thomas} S. Thomas, Phys. Rev. Lett. \textbf{89}, 081301 (2002) .

\bibitem{Hsu} S.D.H. Hsu, Phys. Lett. B \textbf{594}, 13 (2004) .

\bibitem{Li} M. Li, Phys. Lett. B \textbf{603}, 1 (2004).

\bibitem{Odintsov} S. Nojiri and S. D. Odintsov, Gen. Rel. Grav. {\bf 38}, 1285 (2006).

\bibitem{Odintsov1} S. Nojiri and S. D. Odintsov, Eur. Phys. J. C {\bf 77}, 528 (2017).

\bibitem{Odintsov2} S. Nojiri, S. D. Odintsov and E. N. Saridakis, Phys. Lett. B {\bf 797}, 134829 (2019).

\bibitem{Odintsov3} S. Nojiri, S. D. Odintsov and E. N. Saridakis, Nucl. Phys. B {\bf 949}, 114790 (2019).

\bibitem{Guberina} B. Guberina, R. Horvat, H. Nikolic, J. Cosmol. Astropart. Phys. \textbf{01}, 012 (2007).

\bibitem{Myung} Y.S. Myung, Phys. Lett. B \textbf{652}, 223 (2007).

\bibitem{Ghaffari} S. Ghaffari, M.H. Dehghani, A. Sheykhi, Phys. Rev. D \textbf{89}, 123009 (2014).

\bibitem{Zadeh} M.A. Zadeh, A. Sheykhi, H. Moradpour, Int. J. Mod. Phys. D \textbf{26}, 1750080 (2017).

\bibitem{Hao} W. Hao, Commun. Theor. Phys. \textbf{52}, 743 (2009).

\bibitem{Wang} S. Wang, Y. Wang, M. Li, Phys. Rep. \textbf{696}, 1 (2017).

\bibitem{Wang-a} B. Wang, E. Abdalla, F. Atrio-Barandela, D. Pavon, Rep. Prog. Phys. \textbf{79}, 096901 (2016).

\bibitem{Shen} J. Shen, B. Wang, E. Abdalla, R.K. Su, Phys. Lett. B \textbf{609}, 200 (2005).

\bibitem{Sheykhi} A. Sheykhi, Phys. Lett. B \textbf{680}, 113 (2009).

\bibitem{Zhang} X. Zhang, Phys. Rev. D \textbf{74}, 103505 (2006).

\bibitem{Sheykhi-a} A. Sheykhi, et al., Gen. Relativ. Gravit. \textbf{44}, 623 (2012).

\bibitem{Setare} M.R. Setare, M. Jamil, Europhys. Lett. \textbf{92}, 49003 (2010).

\bibitem{Sheykhi-b} A. Sheykhi, Phys. Lett. B \textbf{681}, 205 (2009).

\bibitem{Sheykhi-c} A. Sheykhi, Phys. Lett. B \textbf{682}, 329 (2010).

\bibitem{Karami} K. Karami, M.S. Khaledian, M. Jamil, Phys. Scr. \textbf{83}, 025901 (2011).

\bibitem{Agostino} R. D'Agostino, Phys. Rev. D \textbf{99}, 103524 (2019).


\bibitem{Li-a} M. Li, X. D. Li,S. Wang, X. Zhang, JCAP \textbf{06}, 036 (2009).


\bibitem{Li-b} M. Li, X. D. Li,S. Wang, Y. Wang, X. Zhang, JCAP \textbf{12}, 014 (2009).


\bibitem{saaidi} Kh. Saaidi, A. Aghamohammadi, Phys. Scripta \textbf{83}, 025902 (2011).

\bibitem{saaidi-a} Kh. Saaidi, A. Aghamohammadi, B. Sabe, O. Farooq, Int. J. Mod. Phys. D \textbf{21}, 1250057 (2012).

\bibitem{Aghamohammadi} A. Aghamohammadi, Kh. Saaidi, A. Mohammadi. H. Sheikhahmadi, T. Golanbari, S. W. Rabiei, Astrophys. Spac. Sci. \textbf{345}, 17 (2013).

\bibitem{karami-a} K. karami, A. Sorouri, Phys. Scripta \textbf{82}, 025901 (2010).

\bibitem{karami-b} K. karami, A. Abdolmaleki, Int. J. Theor. Phys. \textbf{50}, 1656 (2011).

\bibitem{karami-c} K. karami, arXiv: 1002.0431v1 (2018).

\bibitem{karami-d} K. karami, A. Abdolmaleki, N. Sahraei, S. Ghaffari, JHEP \textbf{1108}, 150 (2011).


\bibitem{Masi} M. Masi, Phys. Lett. A \textbf{338}, 217 (2005).

\bibitem{Touchette} H. Touchette, Phys. A \textbf{305}, 84 (2002).

\bibitem{Biro} T.S. Biró, P. Ván, Phys. Rev. E \textbf{83}, 061147 (2011).

\bibitem{Tsallis} C. Tsallis, Entropy \textbf{13}, 1765 (2011).

\bibitem{Renyi} A. Rényi, Probability Theory (North-Holland, Amsterdam, 1970).

\bibitem{Tsallis-a} C. Tsallis, J. Stat. Phys. \textbf{52}, 479 (1988).

\bibitem{Odintsov4} S. Nojiri, S. D. Odintsov and E. N. Saridakis, Eur. Phys. J. C {\bf 79}, 242 (2019).

\bibitem{Abe} S. Abe, Phys. Rev. E \textbf{63}, 061105 (2001).

\bibitem{Abe-a} S. Abe, Foundations of Nonextensive Statistical Mechanics, in Chaos, Nonlinearity, Complexity. Studies in Fuzziness and Soft
Computing 206, ed. by A. Sengupta (Springer, Berlin, 2006).

\bibitem{Majhi} A. Majhi, Phys. Lett. B \textbf{775}, 32 (2017).

\bibitem{Biro-a} T.S. Biró, V.G. Czinner, Phys. Lett. B \textbf{726}, 861 (2013).

\bibitem{Sayahian}  A. Sayahian Jahromi et al., Phys. Lett. B \textbf{780}, 21 (2018).


\bibitem{Komatsu} N. Komatsu, Eur. Phys. J. C \textbf{77}, 229 (2017).

\bibitem{Moradpour} H. Moradpour, A. Bonilla, E. M. C. Abreu, J. A. Neto, Phys. Rev. D \textbf{96}, 123504 (2017).

\bibitem{Moradpour-a} H. Moradpour, A. Sheykhi, C. Corda, I.G. Salako, Phys. Lett. B \textbf{783}, 82 (2018).

\bibitem{Moradpour-b} H. Moradpour, Int. J. Theor. Phys. \textbf{55}, 4176 (2016).

\bibitem{Moradpour-c} H. Moradpour, et. al. Eur. Phys. J. C \textbf{78} 829 (2018).

\bibitem{Abreu} E.M.C. Abreu, J. AnaniasNeto, A.C.R. Mendes,W. Oliveira, Phys. A \textbf{392}, 5154 (2013).

\bibitem{Abreu-a} E.M.C. Abreu, J. Ananias Neto. Phys. Lett. B \textbf{727}, 524 (2013).

\bibitem{Barboza} E.M. Barboza Jr., R.C. Nunes, E.M.C. Abreu, J.A. Neto, Phys. A Stat. Mech. Appl. \textbf{436}, 301 (2015).

\bibitem{Nunes} R.C. Nunes et al., JCAP \textbf{08}, 051 (2016).


\bibitem{Bialas} A. Bialas, W. Czyz, EPL \textbf{83}, 60009 (2008).

\bibitem{Czinner} V.G. Czinner, H. Iguchi, Phys. Lett. B \textbf{752}, 306 (2016).


\bibitem{Poisson} E. Poisson, W. Israel, Phys. Rev. D \textbf{41}, 1796 (1990).

\bibitem{Hayward} S.A. Hayward, Phys. Rev. D \textbf{53}, 1938 (1996).

\bibitem{Gong} Y. G. Gong, A. Wang, Phys. Rev. Lett. \textbf{99}, 211301 (2007).

\bibitem{Jacobson} T. Jacobson, Phys. Rev. Lett. \textbf{75}, 1260 (1995).

\bibitem{Cai-a} R. G. Cai, S.P. Kim, JHEP \textbf{02}, 050 (2005).

\bibitem{Cai-b} R. G. Cai, L.M. Cao, Y. P. Hu, JHEP \textbf{08}, 090 (2008).

\bibitem{Planck2018} Planck Collaboration: N. Aghanim et al. arXiv:1807.0620 (2019).

\bibitem{Miao-Li} Miao Li et al, JCAP \textbf{06}, 036 (2009); Miao Li et al, JCAP \textbf{12}, 014 (2009).



\end{thebibliography}
\end{document}